\documentclass[12pt]{article}
\usepackage{amsmath,amsthm}
\usepackage{amssymb}
\usepackage{extarrows}
\usepackage{amsfonts}
\usepackage{graphicx}
\usepackage{xcolor}
\usepackage{lineno,hyperref}
\usepackage[capitalize]{cleveref}
\usepackage{enumerate}
\usepackage{subfigure}
\usepackage{float}
\usepackage{fullpage}
\usepackage{authblk}

\theoremstyle{plain}
\newtheorem{thm}{Theorem}[section]

\theoremstyle{definition}

\newtheorem{rem}[thm]{Remark}

\numberwithin{equation}{section}
\newtheorem{example}{Example}

\begin{document}

\title{\vspace{-70pt} Opinion Dynamics over Migration Networks}

% \author{L\H{o}rinc M\'arton$^{2,1}$, Stefanie Winkelmann$^{1}$,  Mauricio J. del Razo$^{1,3}$, \\
% Nata\v sa Djurdjevac Conrad$^{1}$\\ \\
% $\ ^{1}$Zuse-Institute Berlin \\
% $\ ^{2}$Sapientia Hungarian University of Transylvania \\
% $\ ^{3}$Freie Universität Berlin}

\author[1]{L\H{o}rinc M\'arton}
\author[2]{Stefanie Winkelmann}
\author[2,3]{Mauricio J. del Razo}
\author[2]{Nata\v sa Djurdjevac Conrad\thanks{Corresponding author: natasa.conrad@zib.de}}

\affil[1]{Sapientia Hungarian University of Transylvania, Romania}
\affil[2]{Zuse Institute Berlin, Berlin, Germany}
\affil[3]{Freie Universität Berlin, Berlin, Germany}

%\date{}

\maketitle

\tableofcontents

\pagebreak

\begin{abstract}
%150 to 250 words
Opinions play a crucial role in shaping collective phenomena such as political polarization, cultural integration and demographic change. By continuously changing social environments in which opinions evolve, human migration serves as an important driver of collective opinion formation. While migration and opinion dynamics have both been extensively studied, the few existing models that couple the two are primarily deterministic and therefore cannot capture demographic fluctuations, finite-size effects or stochastic transitions between emergent collective states. 
To address this limitation, we introduce a unifying stochastic framework for opinion dynamics over migration networks that couples local opinion transitions, demographic processes and migration between communities. 
The dynamics are formulated through a spatio--temporal master equation, which provides a probabilistic description of the underlying population process. From this microscopic representation, we derive deterministic mean-field equations governing the co-evolution of community sizes and opinion compositions, thereby linking agent-level interactions to macroscopic population behavior.
Using two representative case studies, we demonstrate how stochasticity and migration can qualitatively change the emergent dynamics and collective outcomes, including the emergence of consensus, polarization and the stabilization of oscillatory opinion dynamics.
These examples highlight the rich interplay between social interactions, demographic change and migration in deterministic and stochastic settings, and they demonstrate that migration should be viewed as an integral component of collective opinion formation rather than only an external demographic process.
\end{abstract}

\section{Introduction}

One of the central challenges in computational social science is understanding how opinions form, evolve and spread through populations, as such collective dynamics shape political behavior, public health decisions, social cohesion and other societal outcomes. However, opinions, attitudes and cultural traits are not distributed uniformly in space, but often have strong spatial heterogeneity. For example, studies have shown that the factors influencing voting behavior differ significantly between geographical regions~\cite{Ziqi_GIS2022,Horan_ASAP2024} and that regional traditions of tolerance toward immigrants can last throughout generations~\cite{Fielding_2018}.  Similar spatial clustering has been observed for environmental attitudes~\cite{Villa_SS2022}, climate policy support~\cite{Goerg_SR2024} and health-related decision acceptance~\cite{Meara_PH2014}. Together, these findings indicate that attitudes and opinions often cluster geographically, shaped by exposure to social environments, local institutions and cultural traditions. 

Migration continuously changes the composition of social clusters by redistributing individuals between communities. In this way, migration can influence opinion dynamics through changes in social interactions, exposure to diversity and cultural transmission. Empirical studies have shown that depending on the type and degree of intergroup contact, migration may either strengthen social divisions or foster social cohesiveness. For example, regular interactions between native and immigrant populations can reduce prejudice and foster more tolerant attitudes~\cite{Bentsen_JIMI2022}, while limited contact may instead strengthen stereotypes and polarization. Migration can also affect the transmission of political values across generations, as immigrants often keep parts of the value systems from their countries of origin, while gradually adapting to the norms of their new communities~\cite{Gonnot_WP2023}. Opinions in the communities of origin can also be impacted by migration, such as when return migrants spread democratic principles~\cite{Ivlevs_2020,Maroc_Springer2019}. Additionally, second-generation immigrants show  heterogeneous political preferences across rural and urban environments~\cite{Moriconi_JUE2025}, and attitudes toward immigration and multiculturalism are often different between migrants and non-migrants~\cite{Sedovic_EJP2025}.  Thus, understanding collective opinion dynamics requires accounting not only for how individuals change their opinions through local interactions, but also for how populations change, move across space and reshape the social environments in which opinions are formed.

A rich variety of mathematical models has been developed to study opinion dynamics and the emergence of collective phenomena, such as consensus formation, polarization and fragmentation. Classical approaches include voter models~\cite{Holley1975}, Sznajd-type models~\cite{Sznajd2000}, bounded-confidence models~\cite{HegselmannKrause2002} and kinetic models ~\cite{Giacomo_KRM2017,Albi_Calzola_Dimarco_2025}. More recently, agent-based models have been employed to investigate the influence of co-evolving social processes~\cite{Conrad_E2022}, social media~\cite{Helfmann_SR2023}, and heterogeneous interaction structures~\cite{zimper2025meanfieldoptimalcontrolstochastic} 
on opinion formation. 
Models incorporating local mobility have further been used to study the interplay between opinion formation and movement of individuals in continuous space~\cite{Schweitzer2000}.
However, despite their diversity, most opinion-dynamics frameworks assume fixed populations and do not account for demographic processes 
or migration between communities that change population sizes and compositions over time. 

In parallel, migration has been extensively studied using compartmental and metapopulation models that have been successfully applied in ecology, epidemiology and population dynamics~\cite{Barbosa_PR2018,Haddad_2010}. These models typically describe populations distributed across multiple spatial regions that are connected through migration flows, e.g. migration-driven population models~\cite{MARTON_TPB2022}, migration-connected Lotka--Volterra systems~\cite{Marton_NLD2025} and spatio--temporal compartmental models for epidemic spreading~\cite{WinkelmannZonkerSchuetteetal.2021}. While these frameworks capture demographic and mobility processes, they generally do not consider opinion dynamics and social influence mechanisms.

A few recent studies have introduced frameworks that simultaneously model opinion dynamics, demographic processes and migration between communities. 
For instance, a deterministic metapopulation model has demonstrated that migration significantly affects the evolution and spatial distribution of ideological groups~\cite{Vitanov_ACS2012}. Similarly, diffusion-driven pattern formation in network-based opinion models has revealed that migration can preserve opinion diversity through spatial self-organization~\cite{MauchGross2026}. Nevertheless, these approaches are deterministic and therefore cannot account for demographic fluctuations, finite-size effects, or stochastic transitions between different collective behaviors observed in real-world social systems.

To bridge this gap, we introduce a unifying stochastic framework that couples opinion dynamics, demographic processes and migration on networks. We consider a collection of communities connected by migration flows, where individuals may change opinions through local opinion transitions, enter or leave communities through demographic processes and migrate between communities. From a mathematical perspective, the resulting dynamics can be viewed as an analogue of biochemical reaction--diffusion systems. Within this analogy, local opinion transitions correspond to reactions and migration acts as a diffusion mechanism between communities. This correspondence enables the transfer of analytical and numerical techniques from reaction--diffusion theory~\cite{erban2020stochastic,isaacson2013convergent,winkelmann2016spatiotemporal,Winkelmann_2020} to the study of opinion dynamics.

We formulate the system dynamics through a spatio--temporal master equation, providing a probabilistic description of the underlying stochastic process. From this microscopic representation, we derive deterministic mean-field equations governing both the evolution of community sizes and the evolution of opinion compositions within communities. The framework therefore provides a consistent description across microscopic, mesoscopic, and macroscopic scales, connecting individual-level interactions, stochastic population dynamics and deterministic collective behavior within a unified multiscale framework. In the large-population limit, the mean-field equations emerge as coupled opinion--migration dynamics, analogous to deterministic reaction--diffusion systems~\cite{anderson2015stochastic}. Through this multiscale perspective, our framework offers a flexible and mathematically consistent approach for studying the interplay between opinion formation, demographic change and migration.

The remainder of the paper is organized as follows. In Section~\ref{sec:setting}, we introduce our modeling framework and specify the local opinion transitions, local demographics and migration mechanisms. We formulate the system dynamics in Section~\ref{sec:population_process}, where we first derive the stochastic evolution of the system using a spatio--temporal master equation and then present the corresponding deterministic mean-field approximation. In Section~\ref{sec:case_stdy}, we demonstrate the applicability of our model using two representative case studies. Finally, in Section~\ref{sec:conclusion}, we provide a brief conclusion, discuss the limitations of the framework and outline potential directions for future research.

\section{Model setup} \label{sec:setting} 

We consider a system of interacting agents. Each agent belongs to exactly one \emph{community} $C_i\in \mathcal{C}$ from a finite set of communities,
\begin{equation}
    \mathcal{C}= \left\{ C_1, ~ C_2, ~ \ldots C_M \right\}, \quad M\in \mathbb{N}.
\end{equation}
Communities can correspond to concrete geographical regions, such as cities or countries, or to abstract social structures, such as political parties or ideological categories. Additionally, each agent holds exactly one \emph{opinion} $O_k\in \mathcal{O}$ from a finite set of opinions,
\begin{equation}\label{eq:Oset}
	\mathcal{O} = \left\{ O_1, ~ O_2, ~ \ldots O_K \right\}, \quad K\in \mathbb{N}.
\end{equation}

Rather than tracking each individual agent in an agent-based description, we adopt a macroscopic perspective and describe the system by the numbers $n_k^{(i)} \in \mathbb{N}_0$ of agents in community $C_i$ holding opinion $O_k$. 
The state of the system is thus represented by the matrix of agent counts
\begin{equation}
    \boldsymbol{n}
    = (n_k^{(i)})_{i=1,\ldots,M;\, k=1,\ldots,K}
    \in \mathbb{N}_0^{K \times M}.
\end{equation}

\begin{figure}
    \centering
    \includegraphics[width=0.8\linewidth]{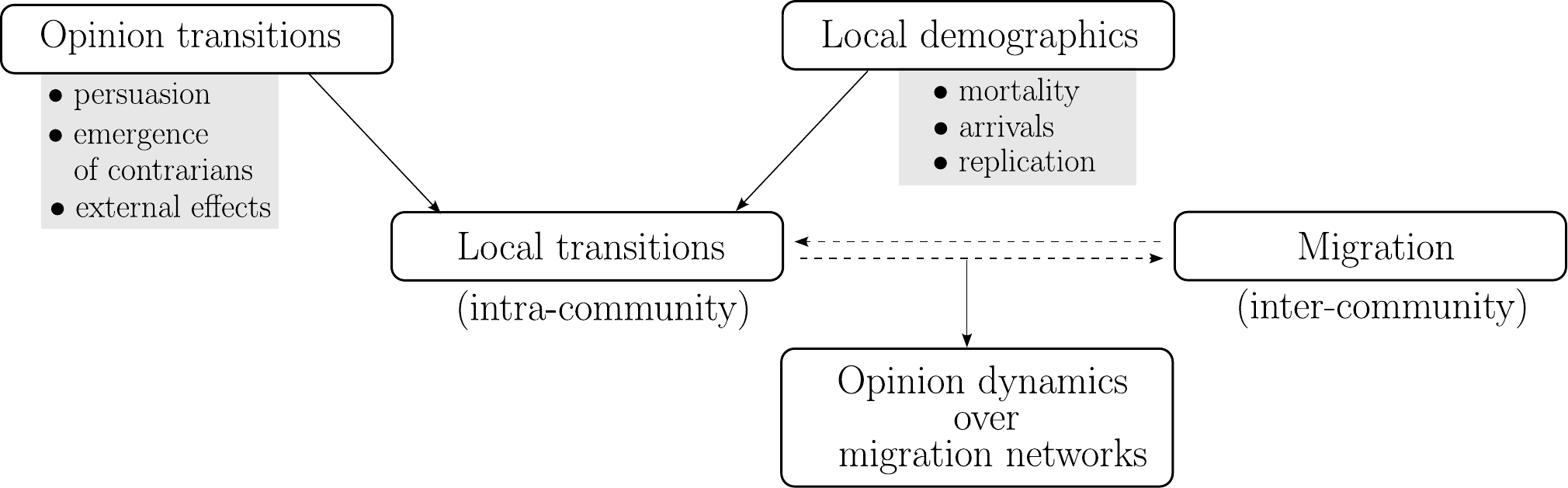}
    \caption{Components of the system dynamics.}
    \label{fig:Model_Structure_alt}
\end{figure}

The temporal evolution of the system is governed by two classes of mechanisms:
\begin{enumerate}
    \item \textbf{Local transitions}, occurring within each community and consisting of:
    \begin{itemize}
        \item \textit{Opinion transitions}, describing changes in individual opinions due to social interactions or external influences (e.g., media exposure or political factors).
        \item \textit{Local demographics}, accounting for changes in community size through inflow and outflow of agents (e.g. births and deaths).
    \end{itemize}
    \item \textbf{Migration}, describing the redistribution of agents between communities due to their mobility.
\end{enumerate}
Local opinion updates and demographic events occur exclusively within communities, whereas communities are coupled only through migration flows. The overall model structure is illustrated in~\cref{fig:Model_Structure_alt}. In the following, we describe these mechanisms in detail.

\subsection{Local transitions} Local transitions are given by discrete events of the form $\mathcal{T}_r$, which take place \emph{within} the communities and can be expressed by stoichiometric equations of the form
\begin{equation}\label{eq:interaction}
    \mathcal{T}_r: \quad \sum_{k=1}^K a_{rk}O_k \; \longrightarrow \; \sum_{k=1}^K b_{rk}O_k, \qquad r=1,...,R .
\end{equation}
Here, $a_{rk},b_{rk}\in \mathbb{N}_0$ are stoichiometric coefficients specifying the number of agents of type $k$ that are entering a transition $\mathcal{T}_r$ and that are produced by this transition, respectively.  
This formalism follows the classical theory of stochastic biochemical reaction networks~\cite{van1992stochastic,kurtz1972relationship}. In this context, the transitions can be of different \textit{order}, depending on the number of agents $\sum_k a_{rk}$ entering the transition. A transition is said to be \textit{zero-order} if it occurs independently of agents and their interactions, \textit{first-order} if it involves a single agent, \textit{second-order} if it involves pairwise interactions between two agents, etc.

A local transition induces a change in the system state of the form
\begin{equation} \label{jump_local}
    \boldsymbol{n} \mapsto \boldsymbol{n} + \boldsymbol{\nu}_r \boldsymbol{e}^\top_i
\end{equation}
where $\boldsymbol{\nu}_r=(\nu_{r1},\ldots,\nu_{rK})^\top\in \mathbb{Z}^K$ is the transition vector defined by
\begin{equation}
    \nu_{rk} = b_{rk} - a_{rk},
\end{equation}
giving the net change in opinion counts due to the transition $\mathcal{T}_r$, and $\boldsymbol{e}_i$ is the indicator vector for the community where the transition occurs. Thus, $\boldsymbol{\nu}_r \boldsymbol{e}^\top_i\in \mathbb{Z}^{K,M}$ is a matrix whose $i$th column equals $\boldsymbol{\nu}_r$ while all other entries are zero.

We distinguish between two classes of local transitions: \emph{opinion transitions}, which change the opinion composition without affecting the population size, and \emph{local demographic transitions}, which modify the total number of agents. 
Accordingly, we partition the set of local transitions into $R_1$ opinion transitions and $R_2$ demographic transitions, with $R=R_1+R_2$. Without loss of generality, we index the opinion transitions by $r=1,\dots,R_1$ and the local demographic transitions by $r=R_1+1,\dots,R$.

\paragraph{Opinion transitions.} 
Agents may change their opinions either through interactions with other agents or as a result of external influences (e.g. mass media or influencers) without direct agent interaction. These transitions do not change the total population size, so the stoichiometric coefficients of the corresponding transitions satisfy:
\begin{equation} \label{eq:coef_bal}
	    \sum_{k=1}^K \nu_{rk}=0,
    \qquad r=1,\ldots,R_1.
\end{equation}
Typical examples of opinion formation are:
\begin{itemize}
	\item \emph{Persuasion}: Two agents with opinions $O_k$ and $O_l$ interact, and the agent with opinion $O_k$ persuades the agent with opinion $O_l$ to adopt its view. This pairwise interaction corresponds to the second-order transition
	\begin{equation}\label{eq:persuaastion}
		O_k + O_l \rightarrow 2 O_k .
	\end{equation}

	\item \emph{Emergence of contrarians}: Dense groups of agents sharing the same opinion can create opportunities for change or generate disappointment. When  two agents with the same opinion $O_k$ interact, one of the agents may change its opinion to a different opinion $O_l$. This mechanism is described by the second-order transition
	\begin{equation}
		2 O_k \rightarrow O_k + O_l.
	\end{equation}

	\item \emph{External effects}: Under external influences, agents with opinion $O_k$ may independently switch to another opinion $O_l$ in a first-order transition
	\begin{equation}\label{eq:ext_effect}
		O_k \rightarrow O_l.
	\end{equation}
   
\end{itemize}

\paragraph{Local demographics.} Like opinion transitions, local demographic transitions involve agents from one community. However, in contrast to opinion transitions, these demographic events do not conserve the total population size, i.e., 
\begin{equation} \label{eq:no_coef_bal}
      \sum_{k=1}^K \nu_{rk}\neq 0,
    \qquad r=R_1+1,\ldots,R.
\end{equation}
Examples of local demographic transitions are:
\begin{itemize}
    \item \emph{Mortality and exiting the migration network:} An agent of type $O_k$ is removed from the community through the first-order transition
    \begin{equation}
        O_k \rightarrow \emptyset.
    \end{equation}
    In this case, no agents are produced by the transition, i.e., $b_{rk} = 0$ for all $k$. 
    \item \emph{Arrivals from outside the migration network:} The entry of a new agent of type $O_k$ into the community is modeled as a zero-order transition
    \begin{equation}
        \emptyset \rightarrow O_k.
    \end{equation}
    In this case, we have $a_{rk}=0$ for all $k$.
    \item \emph{Replication:} An agent of type $O_k$ generates an additional agent with the same opinion, e.g., modeling offspring taking its parent's opinion. This event is described by a first-order transition
    \begin{equation}
        O_k \rightarrow 2 O_k.
    \end{equation}
\end{itemize}

\paragraph{Propensity functions for local transitions.}

For each local transition $\mathcal{T}_r$ and each community $C_i$, we define a \emph{propensity function} $\varphi_r^{(i)}$ which specifies the rate at which the transition occurs given the current system state $\boldsymbol{n}$. We assume that this rate only depends on the local configuration $\boldsymbol{n}^{(i)}=(n_k^{(i)})_{k=1,\ldots K}$ within community $C_i$, while being independent of the other communities. Typically, 
\begin{equation}\label{eq:propensities}
    \varphi_r^{(i)}(\boldsymbol{n}) = \gamma_r^{(i)} \, h_r(\boldsymbol{n}^{(i)}),
\end{equation}
where $\gamma_r^{(i)}\geq 0$ is the rate constant and $h_r:\mathbb{N}_0^K \to [0,\infty)$ is a combinatorial factor counting the number of distinct agent combinations that can participate in transition $\mathcal{T}_r$.

\begin{example}[Propensities for local transitions]\label{ex:mass-action}
 Under mass--action assumptions~\cite{gillespie1977exact} (i.e., assuming well-mixed interactions so that encounter rates are proportional to the product of the numbers of agents involved), exemplary propensity functions take the form:
\begin{itemize}
    \item $\varphi_r^{(i)}(\boldsymbol{n})=\gamma_r^{(i)}n_k^{(i)}n^{(i)}_l$ for an opinion adaptation of the form $O_k + O_l \to 2O_k$,
    \item $\varphi_r^{(i)}(\boldsymbol{n})=\gamma_r^{(i)}n_k^{(i)}$ for external opinion effects $O_k \to O_l$ or outflow $O_k \to \emptyset$,
    \item $\varphi_r^{(i)}(\boldsymbol{n})=\gamma_r^{(i)}$ for inflow/arrivals $\emptyset \to O_k$. 
\end{itemize} 
\end{example}

\subsection{Migration} Agents can move between communities, and the resulting flow of agents interconnects the communities into a migration network. This network can be represented as a directed graph, where every node $i$ represents one community $C_i\in\mathcal{C}$, and there is a directed edge from node $i$ to node $j$ if migration can occur from community $C_i$ to community $C_j$, described by a first-order transition  
\begin{equation}
  \mathcal{M}^{i\to j}: \quad   C_i \to C_j.
\end{equation}
During migration, the opinions of agents are preserved, i.e., individuals carrying opinion $O_k$ in community $C_i$ move to community $C_j$ while retaining the same opinion. Communities can represent either geographical entities (e.g. cities, regions, countries) or social structures in an abstract social space, such as political parties or online communities. Thus, migration can capture both physical relocation of agents and changes in social affiliation. 

A migration event $\mathcal{M}^{i\to j}$ induces a change in the system state of the form
\begin{equation} \label{jump_migration}
    \boldsymbol{n} \mapsto \boldsymbol{n} + \boldsymbol{1}_k^{(i)} - \boldsymbol{1}^{(j)}_k, \quad i\neq j
\end{equation}
for some opinion index $k\in\{1,\dots,K\}$,
where $\boldsymbol{1}_k^{(i)}\in \mathbb{Z}_0^{K,M}$ is a matrix with a single nonzero entry at position $(k,i)$. 

\paragraph{Propensity functions for migration.}

Analogously to local transitions, we associate each possible migration event with a propensity function $\psi_k^{i\to j}$ specifying the rate at which an agent of opinion $O_k$ migrates from community $C_i$ to $C_j$, given the current system state $\boldsymbol{n}$.  
We introduce the migration propensity function as 
\begin{equation} \label{def:psi}
    \psi_k^{i\to j}(\boldsymbol{n}) = p_k^{i\to j}\big(\boldsymbol{n}\big) \,  \mu_k^{(i)}(\boldsymbol{n}) \, n_k^{(i)},
\end{equation}   
where $\mu_k^{(i)}(\boldsymbol{n})\geq 0$ defines the per-capita emigration rate for agents of opinion $O_k$ from community $C_i$, and $p_k^{i\to j}\big(\boldsymbol{n}\big)\in [0,1]$ is the conditional probability of going to community $C_j$. These probabilities satisfy $\sum_{j=1}^Mp_k^{i\to j}\big(\boldsymbol{n}\big) =1$ for every $i$, $k$ and $\boldsymbol{n}$.

In this general formulation, the emigration rates and probabilities depend on the state $\boldsymbol{n}$ of the entire system rather than only on local configurations. The reasoning behind this is that the decision to emigrate from a particular community is influenced not only by the conditions within that community itself, but also by the population distributions and prevailing opinions in other communities. Thus, migration events can be influenced by local opinion transitions and conversely, migration affects local transitions by changing the size and composition of communities.  
Our model captures this feedback loop between local transitions and migration, shown with dashed lines in~\cref{fig:Model_Structure_alt}.

\begin{example}[Propensities for migration]\label{ex:migration_specs}
Typical specifications of the migration rate function $\psi_k^{i\to j}$ include:
\begin{itemize}
    \item \emph{Linear \& state-independent.} 
    Assume that both destination choice and emigration intensity are configuration-independent,
    \begin{equation}
        p_k^{i\to j}(\boldsymbol{n}) \equiv p_k^{i\to j}, 
        \qquad 
        \mu_k^{(i)}(\boldsymbol{n}) \equiv \mu_k^{(i)},
        \qquad 
        \sum_{j\neq i} p_k^{i\to j}=1.
    \end{equation}
    Then 
    \begin{equation}\label{ex:linear_migration_rates}
        \psi_k^{i\to j}(\boldsymbol{n}) =
    p_k^{i\to j}\,\mu_k^{(i)}\,n_k^{(i)}
    \end{equation}
    is linear in $n_k^{(i)}$ with constant coefficients.

    \item \emph{State-dependent but opinion-blind.}
    Destination choice depends on the current population sizes, but is independent of the opinion type $k$.
    For instance, using a normalized radiation attractiveness~\cite{Simini_N2012},
    let $n^{(i)}=\sum_{k=1}^K n_k^{(i)}$ and define
    \begin{equation}
        G_{ij}(\boldsymbol{n})
        :=
        \frac{n^{(i)}\,n^{(j)}}
        {\big(n^{(i)}+\sigma_{ij}(\boldsymbol{n})\big)\big(n^{(i)}+n^{(j)}+\sigma_{ij}(\boldsymbol{n})\big)},
    \end{equation}
    where $\sigma_{ij}(\boldsymbol{n})$ denotes the total population in communities closer to $C_i$ than $C_j$ (excluding $i$ and $j$). 
    The total attractiveness over all possible destinations adds up to almost one. More precisely, the discrepancy from exact normalization equals the fraction of the total population concentrated in the origin community itself. Therefore, when the total migration network population is large compared with the population of a single origin community, this discrepancy becomes negligible, allowing the radiation attractiveness to be interpreted as a probability distribution over destinations. However, for finite systems, an exact probability distribution can be obtained by a simple normalization:
    \begin{equation}
        p_k^{i\to j}(\boldsymbol{n}) \equiv p^{i\to j}(\boldsymbol{n}) :=\frac{G_{ij}(\boldsymbol{n})}{\sum_{\ell\neq i} G_{i\ell}(\boldsymbol{n})},
    \end{equation}
     while keeping $\mu_k^{(i)}$ constant or state-dependent.

    \item \emph{Opinion-sensitive.}
    Destination choice depends on the opinion type through an opinion-specific utility.
    A standard \textit{discrete-choice model}~\cite{Ortega_MS2013} is
    \begin{equation}
        p_k^{i\to j}(\boldsymbol{n})
        :=
        \frac{\exp\!\big(w_k^{i\to j}(\boldsymbol{n})\big)}
        {\sum_{\ell\neq i}\exp\!\big(w_k^{i\to \ell}(\boldsymbol{n})\big)},
    \end{equation}
    where $w_k^{i\to j}(\boldsymbol{n})$ encodes the utility of moving from $C_i$ to $C_j$ for agents with opinion $O_k$
    (e.g. depending on economic factors, infrastructure, social ties, or the opinion composition of $C_j$).
    Again, $\mu_k^{(i)}(\boldsymbol{n})$ may be taken constant or state-dependent.
\end{itemize}
\end{example}

\section{The system dynamics} \label{sec:population_process}

Having introduced the fundamental components of the model---namely the admissible state transitions and their associated propensity functions---we now formulate the dynamics of the interacting agent system. We consider both a stochastic description, based on a continuous-time Markov process, and its deterministic mean-field approximation. In both cases, the dynamics are described at the macroscopic level of population counts within communities, without tracking individual agents.

\subsection{Stochastic dynamics}
\label{sec:stochdyn}

In the spirit of stochastic opinion models with discrete states~\cite{castellano2009statistical,banisch2019opinion}, let
\begin{equation}
    N_k^{(i)}(t) \in \mathbb{N}_0
\end{equation}
denote the \textit{random number of agents} with opinion $O_k$ in community $C_i$ at time $t\geq 0$.
We collect these variables into the matrix 
\begin{equation}
    \boldsymbol{N}(t)=(N_k^{(i)}(t))_{i=1,\ldots ,M;\, k=1,\ldots, K}.
\end{equation}
This defines a stochastic process $(\boldsymbol{N}(t))_{t\geq 0}$ with state space $\mathbb{N}_0^{K \times M}$. We model $\boldsymbol{N}(t)$ as a continuous-time Markov jump process on this discrete state space. The process performs jumps corresponding to local transitions and migration events (see \eqref{jump_local} and \eqref{jump_migration}), with exponentially distributed waiting times determined by the propensity functions $\varphi_r^{(i)}$ and $\psi_k^{i\to j}$. 

There are two equivalent ways to characterize the process $\boldsymbol{N}(t)$: a pathwise representation via the random time-change formula and the associated spatio--temporal master equation. Both are briefly described below. 

\paragraph{Pathwise formulation.} The continuous-time Markov jump process $\boldsymbol{N}(t)$ admits a 
\emph{random time-change representation} of the form
\begin{align}
\begin{aligned}
\label{eq:pathwise}
    \boldsymbol{N}(t) = & \boldsymbol{N}(0) + \sum_{i=1}^M \sum_{r=1}
^R \mathcal{U}_r^{(i)}\left(\int_0^t\varphi_r^{(i)}(\boldsymbol{N}(s)) \,\mathrm{d}s\right) \boldsymbol{\nu}_r \boldsymbol{e}_i^\top \\
& + \sum_{i\neq j} \sum_{k=1}^K \mathcal{U}_k^{i\to j}\left(\int_0^t \psi_k^{i\to j}(\boldsymbol{N}(s)) \,\mathrm{d}s\right)\big(\boldsymbol{1}_k^{(i)}-\boldsymbol{1}_k^{(j)}\big),
\end{aligned}
\end{align}
where $R$ is the number of local transitions, see~\cref{eq:interaction}.
The first sum describes the local transitions within communities (including opinion formation and local demographic events), while the second sum represents migration between communities. 
Here, $\mathcal{U}_r^{(i)}$ and $\mathcal{U}_k^{i\to j}$ are independent unit-rate Poisson processes whose arguments are the integrated propensity functions. Thus, each transition type occurs at random time points with state-dependent rates determined by the corresponding propensities. 
The matrices $\boldsymbol{\nu}_r \boldsymbol{e}_i^\top$ and 
$\boldsymbol{1}_k^{(i)}-\boldsymbol{1}_k^{(j)}$ encode the associated state changes.

\paragraph{Spatio--temporal master equation.}
While the pathwise formulation describes individual sample trajectories of the stochastic process, the same dynamics can equivalently be characterized at the level of probability distributions by the corresponding spatio--temporal master equation.
Let 
\begin{equation}
    p(\boldsymbol{n},t):=\mathbb{P}(\boldsymbol{N}(t)=\boldsymbol{n}\,|\,\boldsymbol{N}(0)=\boldsymbol{n}_0)
\end{equation} 
denote the probability of the process $\boldsymbol{N}(t)$ being in state $\boldsymbol{n}$ at time $t\geq 0$, given some initial state $\boldsymbol{N}(0)=\boldsymbol{n}_0$. The time evolution of this probability is governed by the \emph{spatio--temporal master equation  (STME)}:
\begin{align}
    \frac{dp(\boldsymbol{n},t)}{dt}  = & \sum_{i=1}^M \sum_{r=1}^R \left[ \varphi_r^{(i)}(\boldsymbol{n}-\boldsymbol{\nu}_r \boldsymbol{e}_i^\top)\,p(\boldsymbol{n}-\boldsymbol{\nu}_r \boldsymbol{e}_i^\top,t) \, - \, \varphi_r^{(i)}(\boldsymbol{n})\,p(\boldsymbol{n},t)\right] \label{eq:STME} \\
    & + \sum_{i\neq j} \sum_{k=1}^K \left[ \psi_k^{i\to j}\big(\boldsymbol{n}+\boldsymbol{1}_k^{(i)} - \boldsymbol{1}_k^{(j)}\big )\,p\big(\boldsymbol{n}+\boldsymbol{1}_k^{(i)} - \boldsymbol{1}_k^{(j)},t\big) \, - \,  \psi_k^{i\to j}(\boldsymbol{n})\,p(\boldsymbol{n},t) \right] .\nonumber
\end{align}

As usual, $p(\boldsymbol{n},t)$ is defined to vanish for invalid states (i.e. configurations with negative entries), ensuring that the shifted arguments in the gain terms are well-defined.
This equation accounts for all possible gain and loss events that change the system state.

The STME (also known as the \textit{reaction--diffusion master equation})~\cite{gardiner2009stochastic,erban2009stochastic,Winkelmann_2020,winkelmann2016spatiotemporal}
is widely used in biochemical reaction--diffusion systems, where particles move between compartments and interact locally within each compartment.
In the present context, it provides the exact probabilistic description of the interacting agent system and forms the basis for stochastic simulation and deterministic approximations.

\paragraph{Simulation of the stochastic dynamics.}
The STME defines a high-dimensional system of coupled ordinary differential equations for the probability distribution $p(\boldsymbol{n},t)$. 
Since the state space $\mathbb{N}_0^{K\times M}$ grows combinatorially with the number of opinions and communities, analytical solutions are generally infeasible.
Instead, we simulate the associated continuous-time Markov jump process $\boldsymbol{N}(t)$ using Gillespie’s stochastic simulation algorithm (SSA)~\cite{gillespie1977exact,gibson2000efficient}, which generates statistically exact sample paths consistent with the master equation. 

\begin{rem}[Special case: Local opinion dynamics]

When restricting to a single community (i.e., $M=1$) and omitting migration, the second line of Eq.~\eqref{eq:STME} vanishes and the model reduces to
\begin{equation}
     \frac{dp(\boldsymbol{n},t)}{dt} = \sum_{r=1}^R \left[ \varphi_r(\boldsymbol{n}-\boldsymbol{\nu}_r)\,p(\boldsymbol{n}-\boldsymbol{\nu}_r,t) \, - \, \varphi_r(\boldsymbol{n})\,p(\boldsymbol{n},t)\right].
\end{equation}
This equation is the analogue of the standard \textit{chemical master equation}~\cite{Winkelmann_2020} describing well-mixed reaction kinetics. In this case, the system state reduces to a vector $n\in \mathbb{N}_0^K$, representing the opinion counts in a single, well-mixed population.
\end{rem}

\subsection{Deterministic approximation} \label{sec:deterministic}

For large populations, the stochastic dynamics of the opinion counts admit a deterministic approximation in the sense of a law-of-large-numbers limit~\cite{kurtz1972relationship,kurtz1970solutions,van1992stochastic}. In this regime, random fluctuations become negligible relative to population sizes, and the system can be described by a system of ordinary differential equations governing its average behavior.

We derive two complementary deterministic descriptions: first, equations for the expected opinion counts within each community (\cref{sec:mean-field}), and second, equations for the corresponding relative opinion fractions (\cref{sec:fractions}). While the former track absolute population sizes, the latter focus on the internal composition of communities.

\subsubsection{Mean-field dynamics} \label{sec:mean-field}

We introduce deterministic mean-field variables
\begin{equation}
    m_k^{(i)}(t) \in [0,\infty),
\end{equation}
and denote by
\begin{equation}
    \boldsymbol{m}^{(i)}(t)
    =(m_k^{(i)}(t))_{k=1,\ldots,K}
    \in \mathbb{R}_{\ge0}^K
\end{equation}
the vector of opinion counts in community $C_i$. 
Let $\boldsymbol{m}=(\boldsymbol{m}^{(i)})_{i=1,\ldots,M}$ collect all communities.

One can obtain equations for the mean-field dynamics by taking expectations of the spatio--temporal master equation and applying a mean-field closure~\cite{del2024field,isaacson2022mean,kostre2021coupling,smith2018spatially}. Applying this methodology to \cref{eq:STME} yields
\begin{equation}\label{eq:mean-field}
\frac{d}{dt}m_k^{(i)}
=
\sum_{r=1}^R \nu_{rk}\,\varphi_r^{(i)}(\boldsymbol{m})
+
\sum_{j\neq i}
\left[
\psi_k^{j\to i}(\boldsymbol{m})
-
\psi_k^{i\to j}(\boldsymbol{m})
\right],
\end{equation}
for $k=1,\ldots,K$ and $i=1,\ldots,M$. The solution of~\eqref{eq:mean-field}  provides an approximation of the true expectations,
\begin{equation}
m_k^{(i)}(t) \;\approx\; \mathbb{E}\!\big[N_k^{(i)}(t)\big],
\end{equation}
and becomes exact when all transition propensity functions are linear in the state variables. 
For nonlinear transitions, higher-order correlations between agents are neglected by the mean-field closure. 
However, the approximation becomes increasingly accurate for large populations, where stochastic fluctuations are relatively small compared to the deterministic drift.

\begin{example}[Local demographics.]
Assume that there are $R=R_1+R_2$ local transitions, consisting of $R_1\in \mathbb{N}$ opinion transitions (which are not further specified here) and the following $R_2=3$ demographic events that can occur in each community $C_i$:
\begin{itemize}
    \item mortality $O_k \to \emptyset$ at rate $\alpha_k^{(i)}\ge 0$,
    \item arrivals $\emptyset \to O_k$ at rate $\beta_k^{(i)}\ge 0$,
    \item replication $O_k \to 2O_k$ at rate $\gamma_k^{(i)}\ge 0$.
\end{itemize}
In combination with the migration events, we obtain the mean-field equation
\begin{align}
    \frac{d}{dt}m_k^{(i)}
    &= \sum_{r=1}^{R_1} \nu_{rk}\,\varphi_r^{(i)}(\boldsymbol{m}) \; + \;\left[
    \;-\; \alpha_k^{(i)} m_k^{(i)}
    \;+\; \beta_k^{(i)}
    \;+\; \gamma_k^{(i)} m_k^{(i)} \right]
    \nonumber\\
    &\quad + \sum_{j\neq i}
    \Big[
        \psi_k^{j\to i}(\boldsymbol{m})
        -
        \psi_k^{i\to j}(\boldsymbol{m})
    \Big].
\end{align}
\end{example}

\subsubsection{Relative opinion fractions} \label{sec:fractions}

In many applications, the relative composition of opinions within each community is of primary interest. 
We therefore introduce the community-wise opinion fractions
\begin{equation}\label{eq:x_def}
    x_k^{(i)}(t)
    =
    \frac{m_k^{(i)}(t)}{s^{(i)}(t)},
    \qquad
    s^{(i)}(t)
    =
    \sum_{k=1}^K m_k^{(i)}(t),
\end{equation}
which satisfy $\sum_{k=1}^K x_k^{(i)}(t)=1$. 

To obtain closed dynamics for the fractions, we assume a density-dependent scaling of the transition rates. 
That is, the count-based propensities introduced in 
\cref{sec:setting} are rescaled with appropriate powers of the community size, so that the corresponding per-capita rates remain finite when expressed in terms of fractions. This normalization ensures that the resulting fraction dynamics remain well-defined in the large-population limit.

Since local transition propensities depend only on the 
state of the corresponding community, we write
\[
\varphi_r^{(i)}(\boldsymbol m)
=
\varphi_r^{(i)}(\boldsymbol m^{(i)}).
\]
Similarly, migration propensities depend only on the 
source and destination communities,
\[
\psi_k^{i\to j}(\boldsymbol m)
=
\psi_k^{i\to j}
\big(
\boldsymbol m^{(i)},
\boldsymbol m^{(j)}
\big).
\]
We denote the resulting normalized rates by
\begin{align}
    \tilde{\varphi}_r^{(i)}(\boldsymbol{x}^{(i)})
    &=
    \frac{1}{s^{(i)}}
    \varphi_r^{(i)}\big(s^{(i)}\boldsymbol{x}^{(i)}\big), \label{prop:T_r_x}
    \\
 \tilde{\psi}_k^{\,i\to j}(\boldsymbol{x}^{(i)},\boldsymbol{x}^{(j)})
    &=
    \frac{1}{s^{(i)}}
    \psi_k^{i\to j}\big(
        s^{(i)}\boldsymbol{x}^{(i)},
        s^{(j)}\boldsymbol{x}^{(j)}
    \big).
\end{align}

For example, a pairwise interaction with count-based combinatorial factor 
$m_k^{(i)}m_l^{(i)}$ is scaled by $1/s^{(i)}$, so that
\[
\varphi_r^{(i)}(\boldsymbol{m})
=
\frac{\gamma_r^{(i)}}{s^{(i)}}m_k^{(i)}m_l^{(i)}
\quad\Rightarrow\quad
\tilde{\varphi}_r^{(i)}(\boldsymbol{x}^{(i)})
=
\gamma_r^{(i)}x_k^{(i)}x_l^{(i)}.
\]
More generally, an interaction involving $q$ agents 
is scaled by $(s^{(i)})^{-(q-1)}$.

Applying the quotient rule to~\eqref{eq:mean-field} yields
the fraction dynamics
\begin{align}\label{eq:replicator}
    \frac{d}{dt} x_k^{(i)}
    &=
    \sum_{r=1}^R \nu_{rk}\,
    \tilde{\varphi}_r^{(i)}(\boldsymbol{x}^{(i)})
    \nonumber\\
    &\quad
    +
    \sum_{j\neq i}
    \left[
        \frac{s^{(j)}}{s^{(i)}}\,
        \tilde{\psi}_k^{\,j\to i}(\boldsymbol{x}^{(j)},\boldsymbol{x}^{(i)})
        -
        \tilde{\psi}_k^{\,i\to j}(\boldsymbol{x}^{(i)},\boldsymbol{x}^{(j)})
    \right]
    -
    x_k^{(i)}\,\Phi^{(i)}(\boldsymbol{x}),
\end{align}
where
\begin{equation}\label{eq:Phi_def}
    \Phi^{(i)}(\boldsymbol{x})
    =
    \frac{1}{s^{(i)}}\frac{d}{dt}s^{(i)}
\end{equation}
denotes the average per-capita growth rate of community $C_i$.
The subtraction of $x_k^{(i)}\Phi^{(i)}$ ensures the conservation of
$\sum_k x_k^{(i)}=1$.

\paragraph{Replicator structure.}

System~\eqref{eq:replicator} has the structure of a generalized replicator equation~\cite{hofbauer1998evolutionary,Harper2009}. 
The term $\Phi^{(i)}$ plays the role of an average growth rate ensuring normalization of the fractions. 
If local transitions and migration conserve community sizes, then $\Phi^{(i)}\equiv 0$ and the fraction dynamics depend only on the compositions $\boldsymbol{x}^{(i)}$.

\paragraph{Population dynamics.}

Summing~\eqref{eq:mean-field} over all opinions yields 
\begin{equation}\label{eq:s_dynamics}
    \frac{d}{dt} s^{(i)}
    =
    \sum_{r=1}^R
        \Big(\sum_{k=1}^K \nu_{rk}\Big)
        \varphi_r^{(i)}(\boldsymbol{m})
    +
    \sum_{j\neq i}\sum_{k=1}^K
    \Big[
        \psi_k^{j\to i}(\boldsymbol{m})
        -
        \psi_k^{i\to j}(\boldsymbol{m})
    \Big].
\end{equation}
The first term represents local transitions that change the total population size within a community. 
Only transitions for which $\sum_k \nu_{rk}\neq 0$ contribute to this term, corresponding to local demographic events such as arrivals, mortality, or replication (see~\cref{eq:no_coef_bal}). In contrast, opinion transitions satisfy $\sum_k \nu_{rk}=0$ (see~\cref{eq:coef_bal}) and therefore do not affect the total population size. 
Using the indexing introduced above, only demographic transitions $r=R_1+1,\ldots,R$ contribute to population changes, so that the first sum in~\eqref{eq:s_dynamics} can be restricted to these indices. 

To express the population dynamics consistently in terms of fractions, we substitute $m_k^{(i)}=s^{(i)}x_k^{(i)}$ and use the scaled propensities defined in \eqref{prop:T_r_x}. 
For migration rates linear in $m_k^{(i)}$ 
(see~\cref{ex:linear_migration_rates}), 
this yields the closed population dynamics
\begin{equation}\label{eq:pop_dyn}
    \frac{d}{dt} s^{(i)}
    =
    s^{(i)} 
    \sum_{r=R_1+1}^R
        \Big(\sum_{k=1}^K \nu_{rk}\Big)
        \tilde{\varphi}_r^{(i)}
        (\boldsymbol{x}^{(i)})
    +
    \sum_{j\neq i}
        s^{(j)} 
        \sum_{k=1}^K
            p_k^{j\to i}
            \mu_k^{(j)} 
            x_k^{(j)}
    -
        s^{(i)} 
        \sum_{k=1}^K
            \mu_k^{(i)} 
            x_k^{(i)} .
\end{equation}
Hence, the population growth rate depends explicitly on the opinion composition through the fractions $x_k^{(i)}$, showing how changes in the internal opinion distribution influence the migration balance and the long-term demographic dynamics.

\section{Case studies}\label{sec:case_stdy}

In this section, we present two case studies to illustrate the key features of the proposed modeling frameworks and highlight the interplay between opinion, demographics and migration dynamics and how these influence collective behavior. In particular, we emphasize the qualitative differences between stochastic and deterministic models, showing how intrinsic noise can reshape the system's dynamical behavior and lead to outcomes that cannot be captured by the deterministic model. All simulations were carried out in \textsc{Matlab} to analyze the dynamic behavior of interacting and migrating agents in the proposed model. Deterministic trajectories were computed using the standard Runge--Kutta solver \texttt{ode45}, while stochastic realizations were generated using the Gillespie algorithm~\cite{gillespie1977exact}. 

\subsection{Influence of migration on polarization}

The aim of this case study is to investigate how the relative strength of migration and opinion transition rates influence the dynamical patterns of opinion evolution in interconnected communities.

\paragraph{Problem statement. }We consider two communities $C_1$ and $C_2$ with population sizes $s^{(i)}(t)$ ($i=1,2$) at time $t$. 
No demographic transitions are assumed, so the total population size remains constant, i.e.,
\[
s^{(1)}(t) + s^{(2)}(t) = s,
\qquad t \geq 0,
\]
where $s>0$ denotes the total population of the two interconnected communities.

The set of possible opinions is binary,
\[
\mathcal{O} = \{O_Y, O_N\},
\]
representing the opinions ``Yes'' and ``No''. 
Let $n_Y^{(i)}(t)$ and $n_N^{(i)}(t)$ denote the numbers of agents in community $C_i$ holding opinions $O_Y$ and $O_N$, respectively. 
The corresponding opinion fractions are defined by
\[
x_Y^{(i)}(t)=\frac{n_Y^{(i)}(t)}{s^{(i)}(t)},
\qquad
x_N^{(i)}(t)=\frac{n_N^{(i)}(t)}{s^{(i)}(t)},
\]
and we write
\[
\boldsymbol{x}^{(i)}=(x_Y^{(i)},x_N^{(i)}),
\qquad
\boldsymbol{x}=(\boldsymbol{x}^{(1)},\boldsymbol{x}^{(2)}).
\]

Within each community, we consider the following third-order opinion transitions:
\begin{align}
   \mathcal{T}_1: \quad  
   2 O_Y + O_N 
   \xrightarrow{\gamma} 
   3 O_Y, 
   \\
   \mathcal{T}_2: \quad  
   2 O_N + O_Y 
   \xrightarrow{\gamma} 
   3 O_N,
\end{align}
where $\gamma>0$ denotes the transition rate coefficient, assumed to be identical for both transitions to preserve symmetry between the two opinions. 
These transitions represent a persuasion mechanism (similar to the one in~\eqref{eq:persuaastion}) in which a pair of agents holding the same opinion exerts social pressure on another agent to adopt their opinion, leading to conformity within the local interaction group.

Assuming mass-action kinetics as in Example \ref{ex:mass-action}, the (scaled) propensity function of transition $\mathcal{T}_1$ is given by 
\begin{equation}%\label{phi_scaled}
    \varphi_1^{(i)}(\boldsymbol{n})=\frac{\gamma}{(s^{(i)})^2}\, \binom{n^{(i)}_Y}{2}n^{(i)}_N,
\end{equation}
where $s^{(i)}$ is the total number of agents in community $i$. In terms of fractions, we obtain 
\begin{equation}%\label{tilde_phi_2}
\tilde{\varphi}_1^{(i)}(\boldsymbol{x})= \frac{1}{s^{(i)}}\varphi_1^{(i)}(\boldsymbol{n}) = \frac{\gamma}{(s^{(i)})^3}\, \frac{n^{(i)}_Y(n^{(i)}_Y-1)}{2}n^{(i)}_N = \frac{\gamma}{2}\, x^{(i)}_Y\big(x^{(i)}_Y-\frac{1}{s^{(i)}}\big)x^{(i)}_N.
\end{equation}

\begin{figure}
    \centering
    \textbf{a.} \includegraphics[width=0.46\linewidth]{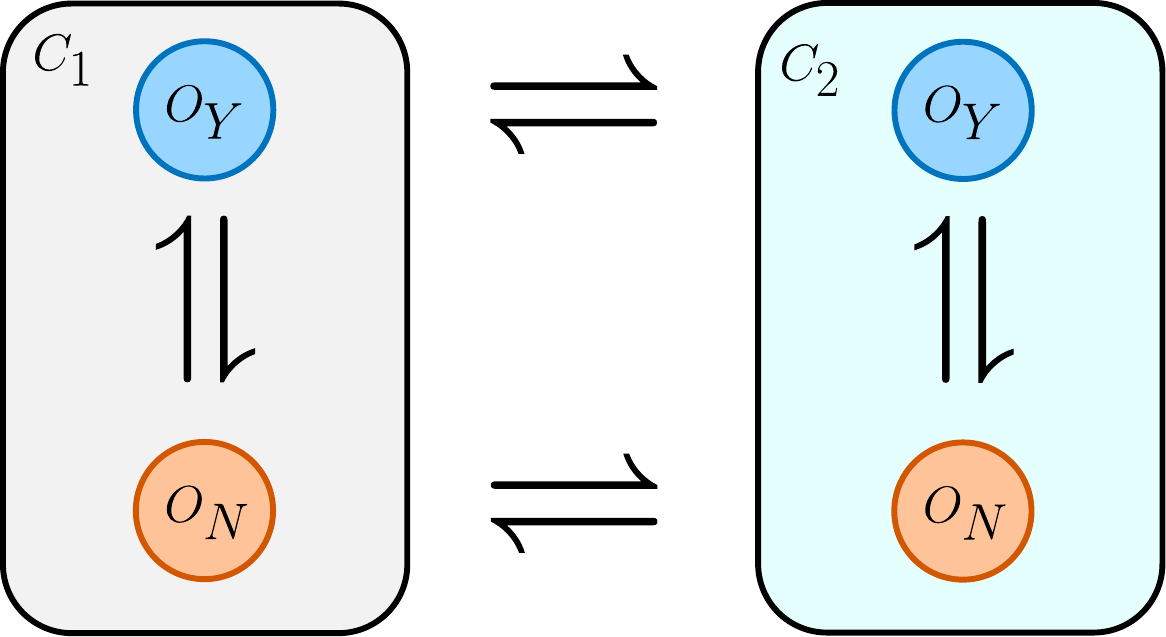}
    \textbf{b.} \includegraphics[width=0.46\linewidth]{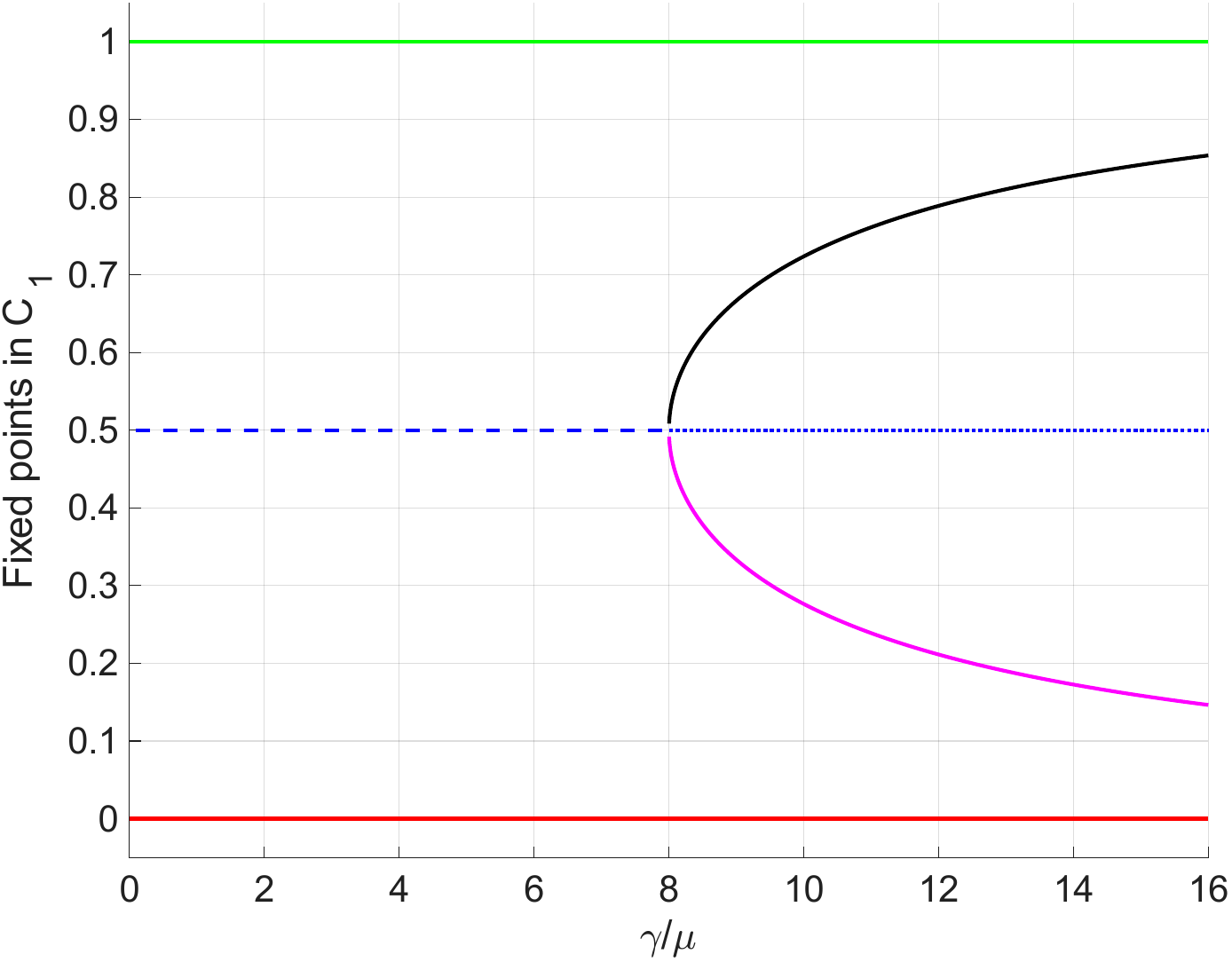}\\[3mm]
    \textbf{c.} \includegraphics[width=0.46\linewidth]{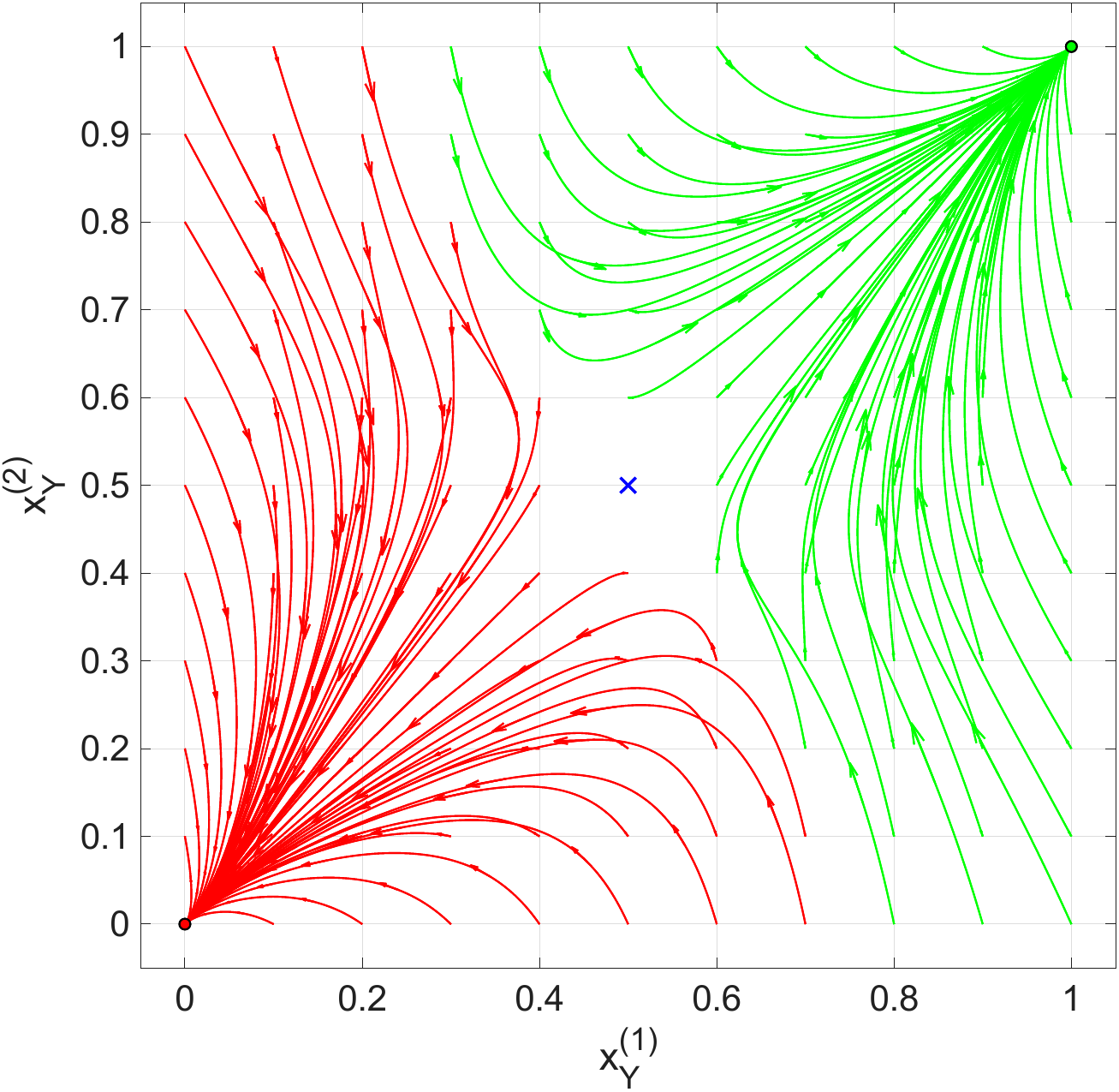}
    \textbf{d.} \includegraphics[width=0.46\linewidth]{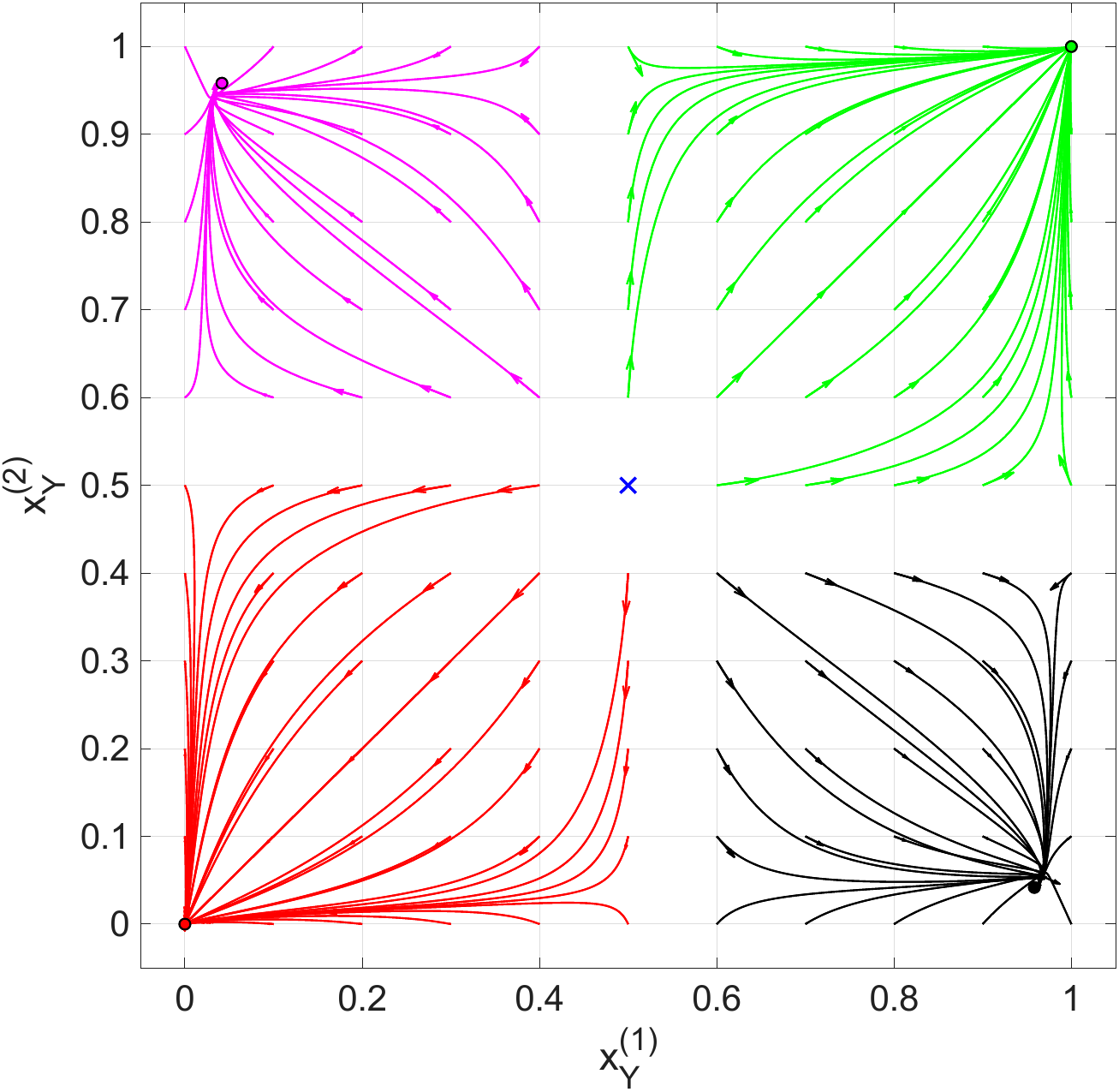}
    \caption{Bifurcation diagram and phase-space analysis.
    \textbf{a.} Schematic illustration of the two-community opinion dynamics model with bidirectional migration and two opinions "Yes" and "No". 
    \textbf{b.} Bifurcation diagram showing the emergence of two additional stable equilibria as the ratio $\gamma/\mu$ increases, in agreement with \cref{eq:eq_stab_mig,eq:eq_unstab,eq:eq_stab_pol}.
    \textbf{c.} Phase portrait in the $(x_Y^{(1)},x_Y^{(2)})$-plane for migration-induced consensus ($\mu=1$, $\gamma=6$). Initial conditions were sampled on the domain $[0,1]\times[0,1]$ with resolution $0.1$. Depending on the initial condition, trajectories converge to one of the two consensus equilibria $(0,0)$ or $(1,1)$. The blue cross marks the unstable equilibrium $(0.5,0.5)$.
    \textbf{d.} Phase portrait for conformity-induced polarization ($\mu=1$, $\gamma=50$). The consensus equilibria remain stable, but their basins of attraction shrink. In addition, two stable polarized equilibria emerge at $(x_Y^{(1)},x_Y^{(2)})=(0.9583,0.0417)$ and $(0.0417,0.9583)$, corresponding to opposite majority opinions in the two communities. Simulations were performed in \textsc{Matlab} using the Runge--Kutta solver \texttt{ode45}.
    } 
    \label{fig:phase_and_bifurcation}
\end{figure}

Since $x_Y^{(i)}(t)+x_N^{(i)}(t)=1$, it is sufficient to track the fraction
\[
    x^{(i)}(t):=x_Y^{(i)}(t),
\]
and then obtain 
\begin{equation}
    \tilde\varphi_1^{(i)}(x^{(i)})
    = \frac{\gamma}{2}\,x^{(i)}
    \Big(x^{(i)}-\frac{1}{s^{(i)}}\Big)
    \big(1-x^{(i)}\big), 
\end{equation}
where we used that $x_N^{(i)}(t)=1-x^{(i)}(t)$. Similarly, we derive that
\begin{equation}
    \tilde\varphi_2^{(i)}(x^{(i)})
    = \frac{\gamma}{2}\,\big(1-x^{(i)}\big)
    \Big(1-x^{(i)}-\frac{1}{s^{(i)}}\Big)
    x^{(i)} .
\end{equation}
In this case study, communities are connected by bi-directional migration flows with state-independent per-capita emigration rates $\mu_Y^{(i)}$ and $\mu_N^{(i)}$ for opinions $O_Y$ and $O_N$, respectively. 
Since only two communities are considered, migration occurs exclusively between them, so that $p^{i\to j}=1$ for $j\neq i$, and the normalized migration propensities are
\begin{equation}
        \tilde{\psi}_Y^{\,i\to j} =
    \mu_Y^{(i)}\,x^{(i)}\,\,\, \text{and} \,\,\,\, \tilde{\psi}_N^{\,i\to j} =
    \mu_N^{(i)}\,(1-x^{(i)})
\end{equation}
for $i\in\{1,2\}$ and $j\neq i$. A schematic illustration of this setup is shown in Figure~\ref{fig:phase_and_bifurcation}a.

Using~\eqref{eq:replicator} and~\eqref{eq:pop_dyn}, the resulting reduced opinion--migration dynamics are
\begin{align}
\frac{d}{dt}x^{(i)}
=&\;
\tilde\varphi_1^{(i)}(x^{(i)})
-
\tilde\varphi_2^{(i)}(x^{(i)})
+
\frac{s^{(j)}}{s^{(i)}}\,\mu_Y^{(j)}x^{(j)}
-
\mu_Y^{(i)}x^{(i)}
-
x^{(i)}\Phi^{(i)}, \\
\Phi^{(i)}
=&\;
\frac{1}{s^{(i)}}\frac{d}{dt}s^{(i)}, \\
\frac{d}{dt}s^{(i)}
=&\;
s^{(j)}
\Big[
\mu_Y^{(j)}x^{(j)}
+
\mu_N^{(j)}\big(1-x^{(j)}\big)
\Big]
-
s^{(i)}
\Big[
\mu_Y^{(i)}x^{(i)}
+
\mu_N^{(i)}\big(1-x^{(i)}\big)
\Big],
\end{align}
for $i\in\{1,2\}$, where $j\neq i$ denotes the other community. Note that the first term for $s^{(i)}$ in equation~\eqref{eq:pop_dyn}  is not present in our formulation, since there are no demographic transitions in this system.

\paragraph{Dynamics without migration.}
If we set the migration rates to zero, i.e., $\mu_Y^{(i)}=\mu_N^{(i)}=0$, $i=1,2$, then we obtain pure local opinion dynamics within each community. 
In this case, the communities evolve independently and the dynamics reduce to
\begin{align*}
\frac{d}{dt} x^{(i)}
&=
\tilde\varphi_1^{(i)}(x^{(i)})
-
\tilde\varphi_2^{(i)}(x^{(i)}) \\
&=
\frac{\gamma}{2} x^{(i)}(1-x^{(i)})
\Big[
(x^{(i)}-\tfrac{1}{s^{(i)}})
-
(1-x^{(i)}-\tfrac{1}{s^{(i)}})
\Big] \\
&=
\frac{\gamma}{2} x^{(i)}(1-x^{(i)})(2x^{(i)}-1) \\
&=
\gamma x^{(i)}(1-x^{(i)})
\big(x^{(i)}-0.5\big).
\end{align*}
Notably, the finite-size correction terms $1/s^{(i)}$ cancel, 
so that the resulting dynamics are independent of the population size. This dynamical system has three fixed points 
$x^{(i)}=0$, $x^{(i)}=0.5$, and $x^{(i)}=1$. 
Linear stability analysis shows that 
$x^{(i)}=0.5$ is unstable, while 
$x^{(i)}=0$ and $x^{(i)}=1$ are stable. 
Consequently, the dynamics exhibit bistability: 
if $x^{(i)}(0)<0.5$, the deterministic system converges to $0$, 
whereas initial values $x^{(i)}(0)>0.5$ lead to convergence to $1$.

\paragraph{Dynamics with migration and numerical analysis.}
To examine the effect of migration on the system dynamics shown in 
\cref{fig:phase_and_bifurcation}a, 
we consider identical migration rates for both opinions, 
i.e.,
\[
\mu_Y^{(i)}=\mu_N^{(i)}=\mu,
\qquad i=1,2.
\]

The system undergoes a pitchfork bifurcation at $\gamma/\mu=8$ (\cref{fig:phase_and_bifurcation}b), so that the number and stability of fixed points depend on this parameter ratio.
For $\gamma/\mu<8$, the system admits three symmetric fixed points 
(\cref{fig:phase_and_bifurcation}c): two stable equilibria,
\begin{equation}\label{eq:eq_stab_mig}
(x_Y^{(1)},x_Y^{(2)})
=(0,0),
\qquad
(x_Y^{(1)},x_Y^{(2)})
=(1,1),
\end{equation}
and one saddle point,
\begin{equation}\label{eq:eq_unstab}
(x_Y^{(1)},x_Y^{(2)})
=
\left(\tfrac{1}{2},\tfrac{1}{2}\right).
\end{equation}

When the parameter ratio crosses the threshold 
$\gamma/\mu=8$, 
the symmetric saddle point loses stability. 
If $\gamma \geq 8\mu$, then, in addition to the symmetric equilibria, the system admits two asymmetric equilibrium states given by
\begin{equation}
\left(
x_Y^{(1)},x_Y^{(2)}
\right)
=
\left(
\frac12+\sqrt{\frac14-\frac{2\mu}{\gamma}},
\frac12-\sqrt{\frac14-\frac{2\mu}{\gamma}}
\right), \label{eq:eq_stab_pol}    
\end{equation}
and
\[
\left(
x_Y^{(1)},x_Y^{(2)}
\right)
=
\left(
\frac12-\sqrt{\frac14-\frac{2\mu}{\gamma}},
\frac12+\sqrt{\frac14-\frac{2\mu}{\gamma}}
\right),
\]
as illustrated in \cref{fig:phase_and_bifurcation}b.

The qualitative system behavior can thus be interpreted in terms of the bifurcation parameter $\gamma/\mu$, which measures the relative strength of local opinion transitions compared to migration:
\begin{itemize}
\item \emph{Migration-induced consensus ($\gamma/\mu \leq 8$).} 

In this regime, migration dominates local reinforcement and suppresses differences between communities. Both communities converge to the same equilibrium opinion, determined by the initial conditions. Unless the initial state lies exactly on the saddle point, the system converges either to $(x_Y^{(1)},x_Y^{(2)})=(0,0)$ or $(x_Y^{(1)},x_Y^{(2)})=(1,1)$, corresponding to global consensus 
(\cref{fig:phase_and_bifurcation}c).

\item \emph{Conformity-induced polarization ($\gamma/\mu > 8$).} In this regime, local reinforcement dominates migration. As a result, symmetry breaking occurs and polarization can emerge between communities. 
Two new stable polarized equilibria appear, in which one community is dominated by opinion $O_Y$ while the other is dominated by $O_N$ 
(\cref{fig:phase_and_bifurcation}d). 
The long-term outcome depends on the initial conditions:
\begin{itemize}
    \item If both $x_Y^{(1)}(0)$ and $x_Y^{(2)}(0)$ are small (respectively large), the system converges to global consensus on opinion $O_N$ (respectively $O_Y$).  
    \item If one initial fraction is small and the other large, the system evolves toward a polarized configuration in which the two communities stabilize with opposite majority opinions.
\end{itemize}
\end{itemize}

Hence, community-level polarization emerges through a symmetry-breaking mechanism: when the reinforcing effect of local opinion transitions exceeds the homogenizing effect of migration, the symmetric equilibrium loses stability. Depending on the initial conditions, the system converges either to global 
consensus or to a stable polarized state.

\begin{figure}
\centering
\textbf{a.}\includegraphics[width=0.3\linewidth]{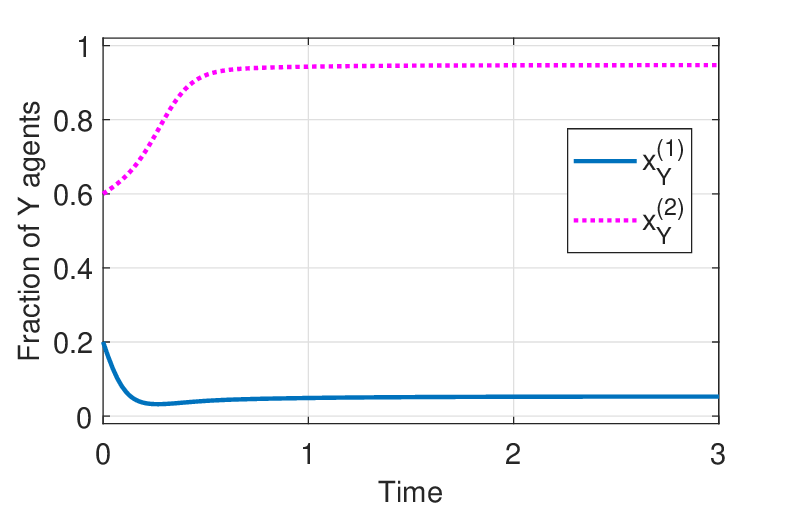}   \textbf{b.}\includegraphics[width=0.3\linewidth]{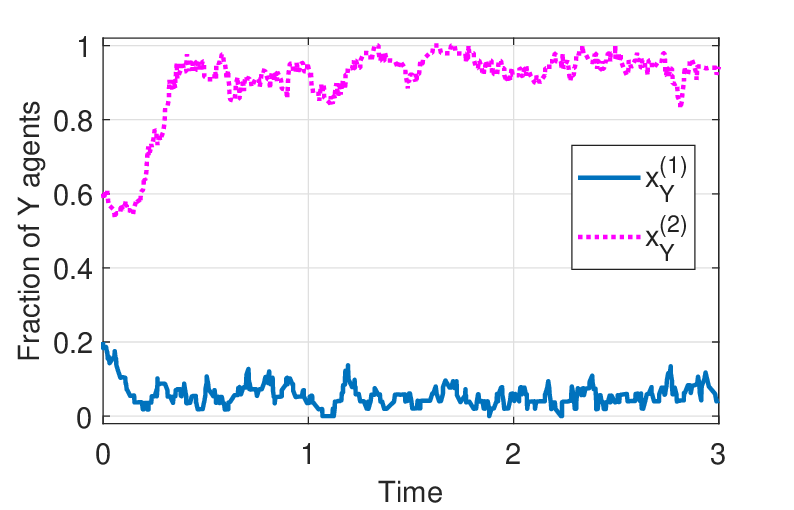}   \textbf{c.}\includegraphics[width=0.3\linewidth]{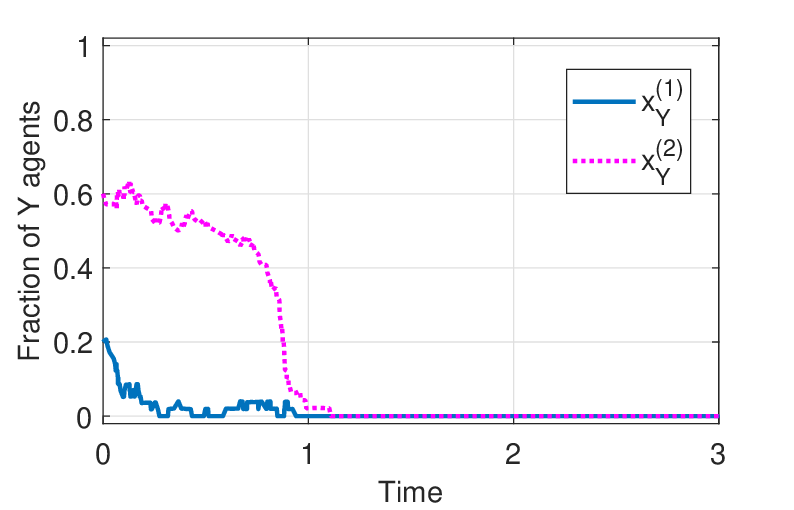}
    \caption{Comparison of deterministic and stochastic dynamics in the conformity-induced polarization regime ($\mu=1$, $\gamma=40$). \textbf{a.} Deterministic solution of \cref{eq:replicator} for the initial condition $x_Y^{(1)} = 0.2$ $x_Y^{(2)} = 0.6$. \textbf{b, c.} Two realizations of the stochastic process \eqref{eq:pathwise} starting from the same initial condition. While the trajectory in \textbf{b} converges to the same equilibrium as the deterministic solution, the trajectory in \textbf{c} converges to a different equilibrium. This illustrates how intrinsic stochastic fluctuations can drive the system into different basins of attraction and thereby alter its long-term outcome.  
    }
    \label{fig:boundaries}
\end{figure}

The behavior of trajectories near the boundaries between the basins of attraction was investigated for both the deterministic (\cref{sec:deterministic}) and stochastic (\cref{sec:stochdyn}) settings. 
For the parameters $\mu=1$, $\gamma=40$ 
(conformity-induced polarization regime) and initial conditions $x_Y^{(1)}=0.2$, $x_Y^{(2)}=0.6$, deterministic simulations indicate convergence to a polarized equilibrium state (see~\cref{fig:boundaries}a). 
However, stochastic simulations of the Markov process 
$\boldsymbol{N}(t)$ defined in~\cref{eq:pathwise} 
(using Gillespie’s algorithm~\cite{gillespie1977exact}) 
reveal a different picture: starting from the same initial condition, trajectories may converge either to a polarized state or to a migration-dominated consensus state due to intrinsic randomness 
(see~\cref{fig:boundaries}b,c). 
This demonstrates that the strict dependence on initial conditions predicted by the deterministic model is relaxed in the stochastic setting, where small random fluctuations can shift the system into different basins of attraction.

To understand when stochastic effects are more likely to occur, we recall that the deterministic model arises from the stochastic process in the large-population limit (\cref{sec:deterministic}). As the community sizes increase, stochastic fluctuations become relatively smaller and the dynamics increasingly resemble those of the deterministic model. 
Conversely, for small communities, fluctuations play a stronger role, and repeated switching between different basins of attraction may occur, even when the deterministic dynamics predict convergence to a single equilibrium. 
These observations highlight that stochastic models capture dynamical features that are not visible in purely deterministic descriptions. A particularly interesting direction for future study is the 
interaction between large and small communities. 
In such settings, hybrid stochastic–deterministic simulation approaches may provide an efficient modeling framework~\cite{Winkelmann_2020}.

\begin{rem}
A large body of work in computational social science has explored polarization arising from nonlinear social interactions, local reinforcement, and network effects. Models ranging from majority-rule dynamics to coevolving network models exhibit stable polarized states when reinforcement dominates mixing or connectivity effects~\cite{Jiazhen_PRL2023, Tomasz_Entropy2022}. This phenomenon is conceptually analogous to the polarization observed in our two-community interaction model.
\end{rem}

\subsection{Influence of migration on opinion oscillations}

Opinion dynamics does not necessarily need to converge monotonically toward a stable consensus but could instead evolve through recurring cycles or oscillations~\cite{Stokes_JMS2024}. Recent theoretical work shows that group-based social influence can generate self-sustained periodic shifts in collective opinions, as demonstrated in models that couple individual and group feedback~\cite{Porter_PRE2025}. Similarly, contrarian voter models subject to time-varying external influence may reproduce oscillatory regimes in which the population’s average opinion tracks the periodic signal~\cite{Gimenez_Entropy2022}. These theoretical insights align, e.g., with political opinion research, which documents substantial vote-share fluctuations in time, indicating that collective political preferences often fluctuate rather than settle into equilibrium~\cite{Fieldhouse_2019}. Such findings suggest that oscillatory dynamics are a natural outcome of how opinions form and evolve in real social systems.

\paragraph{Problem statement.} The underlying mechanisms that exhibit cyclical patterns can be interpreted as the interaction of positive and negative feedback loops operating at both the opinion and community levels. In this study, we investigate a scenario where an opinion $O_k$  becomes dominant within a community $C_i$. Once its local prevalence shows an increasing tendency, it attracts additional agents. However, when the fraction of $O_k$ agents exceeds a crowding threshold or saturates, negative feedback sets in: agents become more likely to change their opinions. As a result, $O_k$ loses its majority position. The process may subsequently repeat with a different opinion. 

\begin{figure}
    \centering    
    \textbf{a.}\hspace{7mm}\includegraphics[width=0.35\linewidth]{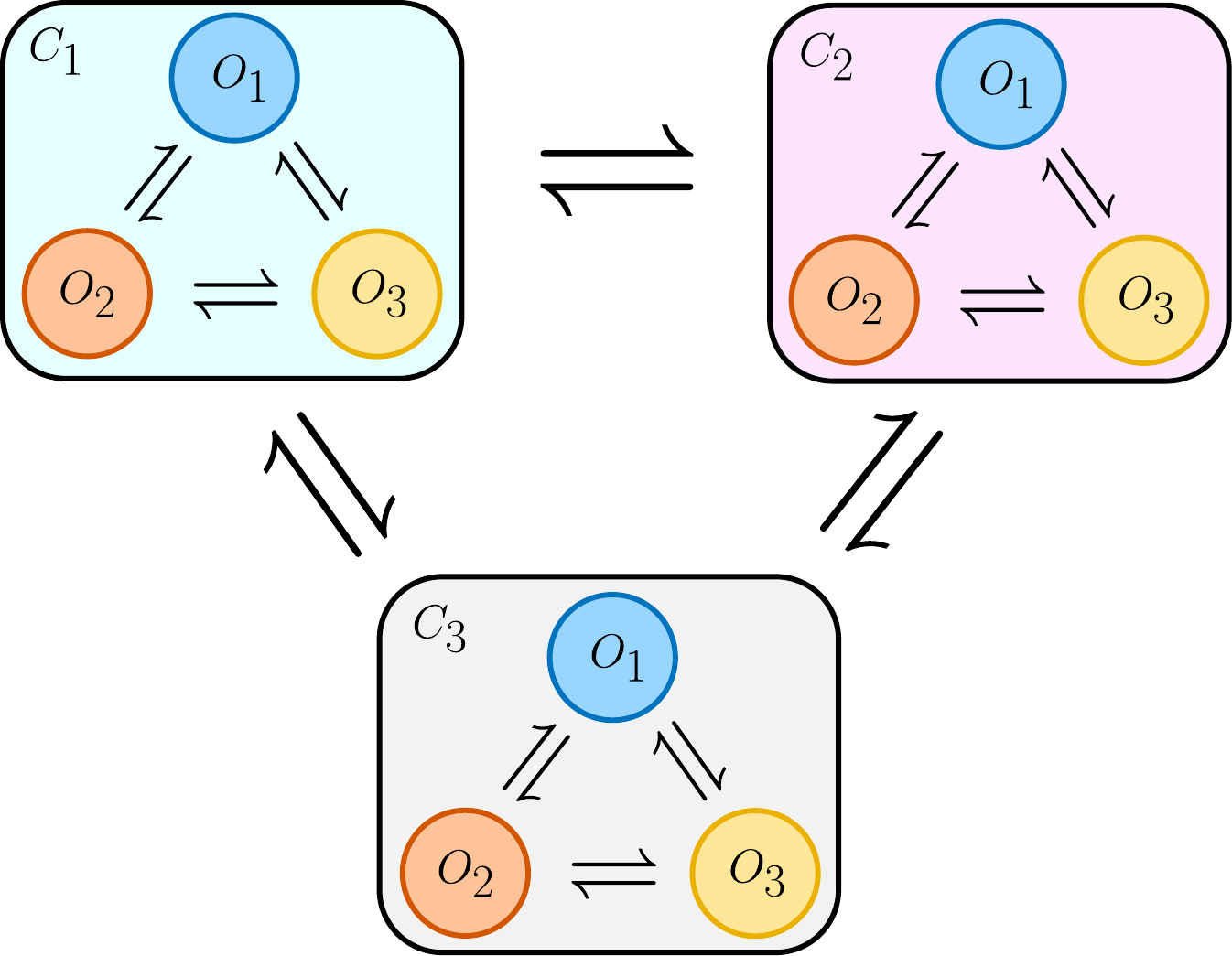} \hspace{7mm} \textbf{b.}\includegraphics[width=0.46\linewidth]{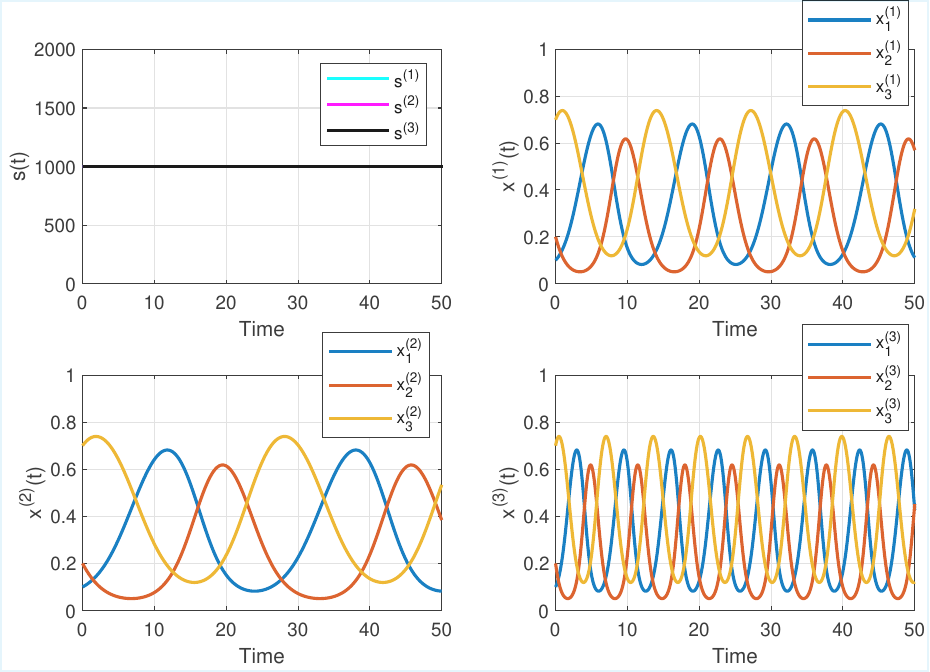} \\ \textbf{c.}\includegraphics[width=0.46\linewidth,  height=170pt]{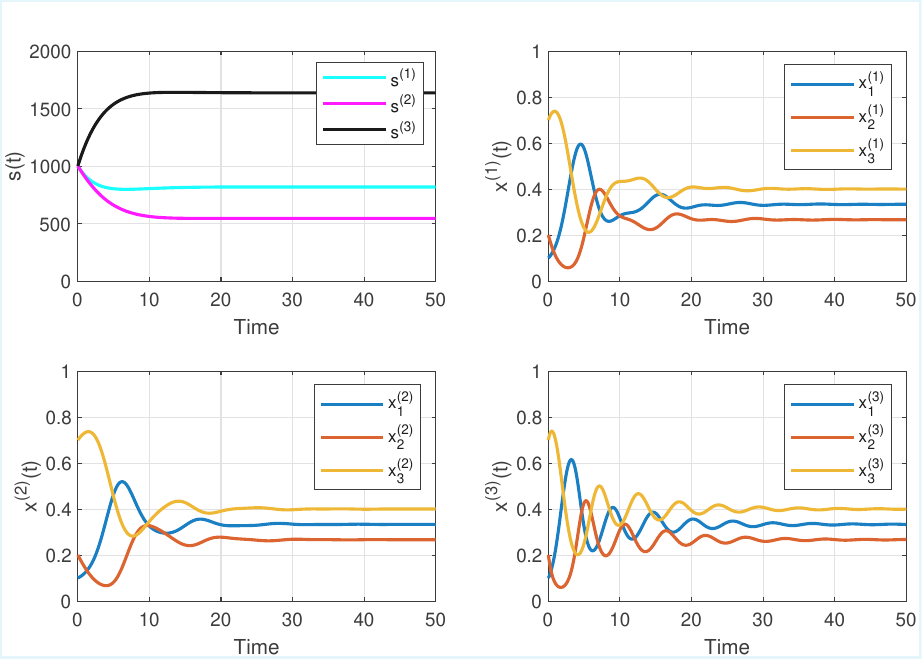}   \textbf{d.}\includegraphics[width=0.46\linewidth, height=170pt]{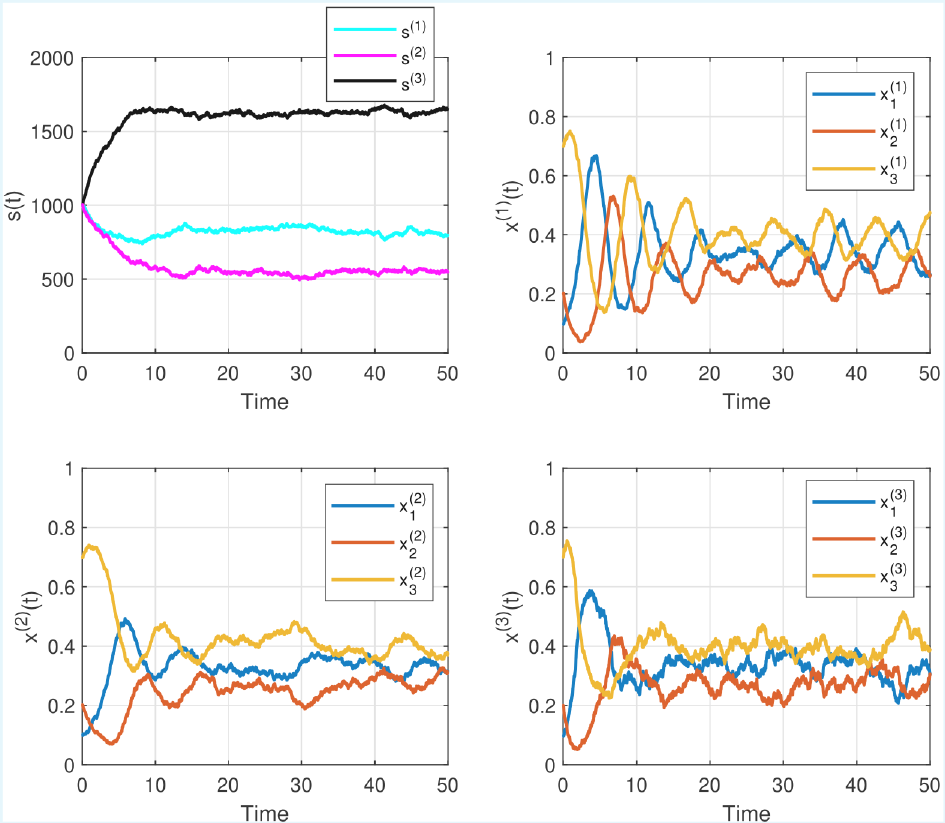}
    \caption{Example with three interconnected communities and three possible opinions. \textbf{a.} Schematic illustration of the opinion dynamics over the migration network. \textbf{b.} Opinion dynamics without migration. Four plots are shown: total population number in each community (top left) and the fraction of agents of the three opinions in community $C_1$ (top right), $C_2$ (bottom left) and $C_3$ (bottom right). \textbf{c.} The same four plots as in \textbf{b} are shown with migration effects turned on. \textbf{d.} The same four plots as in \textbf{b} are shown, but now employing stochastic simulations with the Gillespie algorithm.}
    \label{fig:ring_migration}
\end{figure}

We consider the dynamics of opinions within populations distributed across three communities, where agents can both change opinions through local interactions and migrate between communities. The opinion set is $\left\{ O_1, ~ O_2, ~ O_3\right\}$, and we consider three communities $\{C_1, C_2, C_3\}$ interconnected in a ring, see~\cref{fig:ring_migration}a.

\paragraph{Dynamics without migration.} Within each community, we consider cyclic autocatalytic opinion dynamics
capturing the idea that the abundance of one opinion can increase the likelihood of converting individuals holding other opinions. Specifically, in each community, we consider the following social interactions:
\begin{align}
\begin{aligned}
O_1 + O_2 &\xrightarrow{\gamma_1^{(i)}} 2 O_2, \\
O_2 + O_3 &\xrightarrow{\gamma_2^{(i)}} 2 O_3, \\
O_3 + O_1 &\xrightarrow{\gamma_3^{(i)}} 2 O_1,
\end{aligned}
\end{align}
where the rate constants $\gamma_1^{(i)}, \gamma_2^{(i)}, \gamma_3^{(i)} > 0$ represent the strength of opinion influence. 
The count-based propensities are assumed to be of mass-action type (see Example~\ref{ex:mass-action}),
\[
\varphi_r^{(i)}(\boldsymbol n)
=
\gamma_r^{(i)}
n_k^{(i)}
n_l^{(i)}.
\]
This cyclic structure allows for competition and reinforcement among opinions, analogous to classic \emph{rock-paper-scissor} type chemical kinetics.

\paragraph{Dynamics with migration.} 
A migration event $C_i \to C_j$ from community $C_i$ to community $C_j$ of an agent with opinion $k$ takes place at rate $p_k^{i \rightarrow j}(\boldsymbol{n})\mu_k^{(i)}(\boldsymbol{n})=p_k^{i \rightarrow j}\mu_k^{(i)}$ for a constant $\mu_k^{(i)}>0$ and  
\begin{equation}
    p_k^{i\to j} = \begin{cases}
        1 & \mbox{if } j = i+1\\
        0 & \mbox{else,}
    \end{cases}
\end{equation}
where community indices are interpreted cyclically, i.e.,
$i+1\mapsto 1$ for $i=3$ and $i-1\mapsto 3$ for $i=1$. This leads to a special case of the dynamics illustrated in~\cref{fig:ring_migration}a, namely directed ring dynamics.

In the view of \eqref{eq:replicator} and \eqref{eq:pop_dyn}, the deterministic dynamics of opinion fractions is given by:
\begin{equation}
\frac{d x_k^{(i)}}{dt} = - \gamma_k^{(i)} x_k^{(i)} x_{k+1}^{(i)} + \gamma_{k-1}^{(i)} x_{k-1}^{(i)} x_k^{(i)} + \frac{s^{(i-1)}}{s^{(i)}} \mu_k^{(i-1)}x_k^{(i-1)} - \mu_k^{(i)}x_k^{(i)} - x_k^{(i)} \Phi^{(i)},
\end{equation}
where the $k$-indices are also interpreted cyclically, and where
\begin{align}
    \Phi^{(i)} &=\frac{s^{(i-1)}}{s^{(i)}} \sum_{k=1}^3 \mu_k^{(i-1)} x_k^{(i-1)}  - \sum_{k=1}^3 \mu_k^{(i)} x_k^{(i)}, \\
    \frac{d}{dt} s^{(i)} &= s^{(i-1)} \sum_{k=1}^3 \mu_k^{(i-1)} x_k^{(i-1)}  - s^{(i)} \sum_{k=1}^3 \mu_k^{(i)} x_k^{(i)}.
\end{align}

\paragraph{Numerical experiments.}

Three experimental configurations were examined:
\begin{enumerate}
    \item deterministic simulation without migration,
    \item deterministic simulation with migration, and
    \item stochastic simulation with migration.
\end{enumerate}
In all simulations, the time horizon was set to $T = 50$. The initial population of each community was fixed at $s^{(i)}(0) = 1000$, and the initial distributions of opinions at $t=0$ were set to
\[
x_1^{(i)}(0) = 0.1,\quad 
x_2^{(i)}(0) = 0.2,\quad 
x_3^{(i)}(0) = 0.7,\qquad i = 1,2,3.
\]
The transition coefficients used in the deterministic dynamics were
\begin{align}
\gamma_1^{(1)} = 1.2, \quad 
\gamma_2^{(1)} = 1.0, \quad
\gamma_3^{(1)} = 0.8,  \nonumber \\
\gamma_1^{(2)} = 0.6, \quad 
\gamma_2^{(2)} = 0.5, \quad 
\gamma_3^{(2)} = 0.4, \nonumber \\ 
\gamma_1^{(3)} = 2.4, \quad 
\gamma_2^{(3)} = 2.0, \quad
\gamma_3^{(3)} = 1.6. \nonumber
\end{align}
For simulations that included migration, the emigration rate coefficients were defined as
\[
\mu_k^{(1)} = 0.1,\qquad 
\mu_k^{(2)} = 0.2,\qquad 
\mu_k^{(3)} = 0.3,\qquad k = 1,2,3.
\]

In the first experimental configuration (deterministic without migration), the chosen interaction model (rock–paper–scissors) exhibits neutral cyclic behavior in which competing opinions continuously replace one another without a stable long-term consensus, see~\cref{fig:ring_migration}b. 

In the second experimental configuration, the deterministic dynamics of the three communities are coupled through migration, the exchange of agents introduces an additional mixing mechanism that synchronizes the local opinion dynamics across communities, see~\cref{fig:ring_migration}c.
Instead of each population maintaining closed orbits indefinitely, the exchange of agents gradually aligns their trajectories. The system evolves toward a common fixed point representing a stable coexistence of opinions. Thus, while an isolated system exhibits persistent oscillations, coupling several of them in a feedback structured network could induce collective convergence.

In the third experimental configuration, the ring network of cyclic opinion systems is simulated stochastically using the Gillespie algorithm, the deterministic convergence to the fixed point is perturbed by noise arising from the discrete nature of agents and probabilistic interactions. As a result, instead of settling exactly at the equilibrium, the population fractions fluctuate slightly around the fixed point, see~\cref{fig:ring_migration}d. These fluctuations are typically bounded and do not disrupt the overall stability of the system, but they reflect the natural variability expected in finite populations. Thus, even in a network that promotes deterministic convergence, stochastic effects can maintain persistent, low-amplitude cycles around the otherwise stable coexistence state. For sufficiently small populations, relative fluctuations become larger and may obscure the underlying deterministic structure. Nevertheless, the simulations indicate that the coexistence state remains the dominant organizing feature of the dynamics. In larger networks exhibiting multistability, these stochastic fluctuations could result in switching between stable fixed points, a feature that is not reproducible in the deterministic setting. 

\section{Discussion and Conclusion}\label{sec:conclusion}

In this work, we introduced a general framework for opinion dynamics over migration networks that combines local social interactions and demographic processes with migration between communities. We formulated both stochastic and deterministic descriptions of the resulting dynamics and derived the corresponding spatio--temporal master equation, pathwise representation, and mean-field approximations. The framework is sufficiently flexible to accommodate a wide range of opinion transitions, demographic mechanisms, and migration rules, while naturally coupling changes in opinion composition with changes in community size. The two case studies illustrate how migration can qualitatively alter collective behavior. In the first example, migration promotes consensus, whereas strong local social reinforcement can induce symmetry breaking and polarization between communities. In the second example, isolated communities exhibit persistent oscillations arising from cyclic social interactions, while migration synchronizes the local dynamics and promotes convergence toward a stable coexistence state. Together, these examples demonstrate the versatility of the framework and the diverse dynamical effects that can emerge from the interplay between social interactions and migration,  highlighting migration as a fundamental component of collective opinion formation and not only an external demographic process.

\paragraph{Deterministic and stochastic descriptions.}

A central aspect of the framework is the relationship between deterministic and stochastic models. The deterministic equations arise as large-population approximations of the underlying stochastic dynamics, linking microscopic agent interactions to macroscopic population behavior. The first case study highlights the importance of this distinction. While deterministic trajectories are uniquely determined by their initial conditions, stochastic fluctuations can drive the system into different basins of attraction and thereby alter its long-term outcome. The same example further illustrates how bifurcations can emerge even in relatively simple opinion--migration systems. In contrast, the second case study shows a situation in which stochasticity only produces small fluctuations around a stable coexistence state and does not qualitatively change the dynamics. More generally, larger and more heterogeneous migration networks are expected to exhibit a rich variety of phenomena, including multistability, complex bifurcation structures, stochastic switching between attractors, extinction and growth dynamics, and noise-induced oscillations. Such effects are likely to be particularly relevant in systems containing both large and small populations, where deterministic and stochastic regimes coexist and interact. Their systematic investigation constitutes a promising direction for future research.

\paragraph{Analogy with biochemical reaction--diffusion models and future research.}  

The analogy between opinion--migration systems and biochemical reaction--diffusion systems enables the systematic application of analytical and numerical techniques developed for such systems to the study of social dynamics.
There are, however, several important differences. First, opinion-transition rates and migration rates may depend on the current opinion composition of the involved communities. While this state dependence complicates the analysis, it can be incorporated naturally into the STME framework through suitable propensity functions. Second, whereas diffusion in chemical and biological systems is often modeled as Brownian motion, human migration may exhibit more complex movement patterns, including subdiffusive and superdiffusive behavior such as fractional Brownian motion and Lévy flights~\cite{Barbosa_PR2018,Brockmann_N2006}. Nevertheless, when communities are represented as discrete compartments, many such processes can be approximated through effective Markovian jump rates between communities~\cite{harms2019strong,slkezak2021diffusion}. In this sense, the STME framework remains sufficiently general to describe a broad range of migration mechanisms. Additionally, it opens several avenues for future research, including van Kampen system-size expansions~\cite{van1992stochastic}, large-deviation theory, and Doi--Peliti field-theoretic methods~\cite{del2024field, doi1976second, peliti1985path}. Similarly, hybrid multiscale simulation techniques~\cite{del2025open,kostre2021coupling,del2018grand,smith2018spatially,winkelmann2017hybrid,Winkelmann_2020} can be adapted to opinion--migration models. Exploring these connections is a promising direction for future research, with the potential to transfer theoretical and computational advances from stochastic reaction--diffusion theory to the study of consensus formation, polarization, migration-driven demographic change, and other collective  phenomena in social systems.

\paragraph{Acknowledgment}
This work has been partially funded by the Alexander von Humboldt Foundation and by the Deutsche Forschungsgemeinschaft (DFG, German Research Foundation) under Germany's Excellence Strategy – The Berlin Mathematics Research Center MATH+ (EXC-2046/1, EXC-2046/2, project ID: 390685689) and through the Collaborative Research Center CRC 1114 ``Scaling Cascades in Complex Systems'' (Project No.~235221301).

\bibliographystyle{plain}
\bibliography{Opinion_Migration}

\end{document}